\documentclass[12pt]{article}
\usepackage{aaspp4}		

\def\steve#1{{\bf[Simon, #1 -- Steve]}}

\newcommand{\msun}{\mbox{${\rm M}_\odot$}}

\newcommand{\rsun}{\mbox{${\rm R}_\odot$}}

\newcommand{\mbh}{\mbox{$m_{\rm bh}$}}
\newcommand{\mbar}{\mbox{$\langle m \rangle$}}

\newcommand{\nbody}{\mbox{{{\em N}-body}}}

\newcommand{\rvir}{\mbox{${r_{\rm vir}}$}}

\def\apgt{\ {\raise-.5ex\hbox{$\buildrel>\over\sim$}}\ }
\def\aplt{\ {\raise-.5ex\hbox{$\buildrel<\over\sim$}}\ }

\righthead{Black hole mergers in the universe}
\lefthead{S.\ F.\ Portegies Zwart \& S. L. W. McMillan}

\begin{document}


\title{Black hole mergers in the universe}

\medskip 

\author{Simon F.\ Portegies Zwart$^1$
	\and
	Stephen L.\ W.\ McMillan$^2$
       }


$^1$ Institute for Astrophysical Research	
		  Boston University,
		  725 Commonwealth Ave.,
		  Boston, MA 02215, USA 
		  spz@komodo.bu.edu
\bigskip

$^2$ Dept.\ of Physics,
		  Drexel University, 
                  Philadelphia, PA 19104, USA
                  steve@kepler.physics.drexel.edu
\bigskip

\slugcomment{Simon Portegies Zwart is a Hubble Fellow}

Subject headings: gravitation --- methods: n-body simulations
---stellar dynamics --- binaries (including multiple): close ---
stars: evolution ---globular clusters: general ---

\newpage

\begin{abstract}\noindent
Mergers of black-hole binaries are expected to release large amounts
of energy in the form of gravitational radiation.  However, binary
evolution models predict merger rates too low to be of observational
interest.  In this paper we explore the possibility that black holes
become members of close binaries via dynamical interactions with other
stars in dense stellar systems.  In star clusters, black holes become
the most massive objects within a few tens of millions of years;
dynamical relaxation then causes them to sink to the cluster core,
where they form binaries.  These black-hole binaries become more
tightly bound by superelastic encounters with other cluster members,
and are ultimately ejected from the cluster.  The majority of escaping
black-hole binaries have orbital periods short enough and
eccentricities high enough that the emission of gravitational
radiation causes them to coalesce within a few billion years.  We
predict a black-hole merger rate of about $1.6 \times 10^{-7}$ per
year per cubic megaparsec, implying gravity wave detection rates
substantially greater than the corresponding rates from neutron star
mergers.  For the first generation Laser Interferometer
Gravitational-Wave Observatory (LIGO-I), we expect about one detection
during the first two years of operation.  For its successor LIGO-II,
the rate rises to roughly one detection per day.  The uncertainties in
these numbers are large. Event rates may drop by about an order of
magnitude if the most massive clusters eject their black hole binaries
early in their evolution.
\end{abstract}

\section{Introduction}
The search for gravitational waves will begin in earnest in January
2002, when LIGO-I (Abramovici et al.~1992)\nocite{LIGO} becomes fully
operational (K.~Thorne, private communication).  The appearance of
this new and wholly unexplored observational window challenges
physicists and astronomers to predict detection rates and source
characteristics.  Mergers of neutron-star binaries are widely regarded
as the most promising sources of gravitational radiation, and
estimates of neutron star merger rates (per unit volume) range from
${\cal R} \sim 1.9 \times 10^{-7}\; h^3\, {\rm yr}^{-1}\,{\rm
Mpc}^{-3}$ (Narayan et al.~1991; Phinney 1991; Portegies Zwart \&
Spreeuw 1996) \nocite{Nar91}\nocite{Phi91}\nocite{PZS96}, where $h=
H_0/100\,{\rm km\,s}^{-1}\,{\rm Mpc}^{-1}$, to roughly ten times this
value (Tutukov \& Yungelson 1993; Lipunov et al.,
1997).\nocite{TY93}\nocite{Lip97} However, even with the most
optimistic assumptions, we can expect a LIGO-I detection rate of only
a few neutron star events per millennium.

Inspiral and merger of black-hole binaries are considerably more
energetic events than neutron star mergers, due to the higher masses
of the black holes (Tutukov \& Yungelson 1993; Lipunov et al.~1997).
\nocite{TY93}\nocite{Lip97} Black-hole binaries can result from the
evolution of two stars which are born in a close binary, experience
several phases of mass transfer, and subsequently survive two
supernovae (Tutukov \& Yungelson 1993).\nocite{TY93}
Calculations of event rates from such field binaries depend
sensitively on many unknown parameters and much poorly understood
physics, but the models generally predict a black hole merger rate
${\cal R} \aplt 2 \times 10^{-9}\; h^3\,{\rm yr}^{-1}\,{\rm Mpc}^{-3}$
(Tutukov \& Yungelson 1993; Portegies Zwart \& Yungelson 1998; Bethe
\& Brown 1999),\nocite{TY93}\nocite{PZY98}\nocite{BB99} substantially
lower than the rate for neutron stars.  An alternative possibility,
which we explore here, is that black holes become members of close
binaries not through internal binary evolution, but rather via
dynamical interactions with other stars in a dense stellar system.

\section{Black hole binaries in star clusters}
Black holes are the products of stars with initial masses exceeding
$\sim$20--25\,\msun (Maeder 1992; Portegies Zwart et al.~1997).
\nocite{Maed92}\nocite{PZ++97} A Scalo (1986)\nocite{scalo86} mass
distribution with a lower mass limit of 0.1\,\msun\, and an upper
limit of 100\,\msun\, has 0.071\% of its stars more massive than
20\,\msun, and 0.045\% more massive than 25\,\msun.  A star cluster
containing $N$ stars thus produces $\sim6\times 10^{-4} N$ black
holes.  Known Galactic black holes have masses $\mbh$ between
6\,\msun\, and 18\,\msun\ (Cowlay 1992).\nocite{Cowlay92} For
definiteness, we adopt $\mbh = 10$\,\msun.

\subsection{Binary formation and dynamical evolution}
A black hole is formed in a supernova explosion.  If the progenitor is
a single star (i.e.\ not a member of a binary), the black hole
experiences little or no recoil and remains a member of the parent
cluster (White \& van Paradijs 1996).\nocite{WP96} If the progenitor
is a member of a binary, mass loss during the supernova may eject the
binary from the cluster potential via the Blaauw mechanism (Blaauw
1962),\nocite{bla62} where conservation of momentum causes recoil in a
binary which loses mass impulsively from one component.  We estimate
that no more than $\sim 10$\% of black holes are ejected from the
cluster immediately following their formation.


After $\sim 40$\,Myr the last supernova has occurred, the mean mass of
the cluster stars is $\mbar \sim 0.56$\,\msun\ (Scalo
1986)\nocite{scalo86}, and black holes are by far the most massive
objects in the system.  Mass segregation causes the black holes to
sink to the cluster core in a fraction $\sim\mbar/\mbh$ of the
half-mass relaxation time.  For a typical globular cluster, the
relaxation time is $\sim1$ Gyr; for a young populous cluster, such as
R\,136 (NGC 2070) in the 30 Doradus region of the Large Magellanic
Cloud (Massey \& Hunter 1998),\nocite{MH98} it is $\sim10$ Myr.

By the time of the last supernova, stellar mass loss has also
significantly diminished and the cluster core starts to contract,
enhancing the formation of binaries by three-body interactions.
Single black holes form binaries preferentially with other black holes
(Kulkarni et al.~1992),\nocite{KHM93} while black holes born in
binaries with a lower-mass stellar companion rapidly exchange the
companion for another black hole.  The result in all cases is a
growing black-hole binary population in the cluster core.  Once
formed, the black-hole binaries become more tightly bound through
superelastic encounters with other cluster members (Heggie 1975;
Kulkarni et al.~1992; Sigurdsson \& Hernquist 1993).
\nocite{heggie75}\nocite{KHM93}\nocite{HS93} On average, following
each close binary--single black hole encounter, the binding energy of
the binary increases by about 20\% (Hut et al.\,1992); \nocite{HMR92}
roughly one third of this energy goes into binary recoil, assuming
equal mass stars.  The minimum binding energy of an escaping
black-hole binary may then be estimated as
\begin{equation}
	E_{b,{\rm min}} \sim 36\, W_0\, \frac{\mbh}{\mbar}\, kT\,,
\end{equation}
where $\frac32 kT$ is the mean stellar kinetic energy and $W_0 =
\mbar|\phi_0|/kT$ is the dimensionless central potential of the
cluster (King 1966).\nocite{king66} By the time the black holes are
ejected, $\mbar \sim 0.4\,\msun$.  Taking $W_0\sim5$--10 as a
representative range, we find $E_{b,{\rm min}}\sim5000$--10000 $kT$.

We have tested and refined the above estimates by performing a series
of \nbody\, simulations within the ``Starlab'' software environment
(Portegies Zwart et al.\, 1999: see {\tt
http::/www.sns.ias.edu/$^\sim$starlab}), using the special-purpose
computer GRAPE-4 to speed up the calculations (Makino et
al.\,1997).\nocite{1997ApJ...480..432M} For most (seven) of our
calculations we used 2048 equal-mass stars with 1\% of them ten times
more massive than the average; two calculations were performed with
4096 stars.  One of the 4096-particle runs contained 0.5\% black
holes; the smaller black-hole fraction did not result in significantly
different behavior.  We also tested alternative initial
configurations, starting some models with the black holes in
primordial binaries with other black holes, or in primordial binaries
with lower-mass stars.

The results of our simulations may be summarized as follows.  Of a
total of 204 black holes, 62 ($\sim 30\%$) were ejected from the model
clusters in the form of black-hole binaries.  A total of 124 ($\sim
61\%$) black holes were ejected single, and one escaping black hole
had a low-mass star as a companion.  The remaining 17 ($\sim 8\%$)
black holes were retained by their parent clusters.  The binding
energies $E_b$ of the ejected black-hole binaries ranged from about
$1000\,kT$ to $10000\,kT$ in a distribution more or less flat in $\log
E_b$, consistent with the assumptions made by Hut et al.\ (1992).
\nocite{HMR92} The eccentricities $e$ followed a roughly thermal
distribution [$p(e) \sim 2e$], with high eccentricities slightly
overrepresented.  About half of the black holes were ejected while the
parent cluster still retained more than 90\% ($\aplt 2$ initial
relaxation times) of its birth mass, and $\apgt 90$\% of the black
holes were ejected before the cluster had lost 30\% (between 4 and 10
relaxation times) of its initial mass.
These findings are in good agreement with previous estimates that
black-hole binaries are ejected within a few Gyr, well before core
collapse occurs (Kulkarni et al.~1993; Sigurdsson \& Hernquist
1993).\nocite{KHM93}\nocite{HS93}

We performed additional calculations incorporating a realistic (Scalo)
mass function, the effects of stellar evolution, and the gravitational
influence of the Galaxy.  Our model clusters generally dissolved
rather quickly (within a few hundred Myr) in the Galactic tidal field.
Clusters which dissolved within $\sim 40$\,Myr (before the last
supernova) had no time to eject their black holes.  However, those
that survived beyond this time were generally able to eject at least
one close black-hole binary before dissolution.

Based on these considerations, we conservatively estimate the number
of ejected black-hole binaries to be about $10^{-4} N$ per star
cluster, more or less independent of the cluster lifetime.

\subsection{Characteristics of the binary population}
The energy of an ejected binary and its orbital separation are coupled
to the dynamical characteristics of the star cluster.  For a cluster
in virial equilibrium, we have $kT \equiv 2E_{\rm kin}/3N =
-E_{\rm pot}/3N = G M^2 / 6 N \rvir$,
where $M$ is the total cluster mass and $\rvir$ is the virial radius.
A black-hole binary with semi-major axis $a$ has $E_b = G m_{\rm bh}^2
/ 2 a$,
so
\begin{equation}
	{E_b \over kT} = 3 N \left( {m_{\rm bh} \over M} \right)^2 
                             {\rvir \over a}\,,
\label{Eq:Ebhbh}\end{equation}
establishing the connection between $a$ and the bulk parameters of the
cluster.

In computing the properties of the black-hole binaries resulting from
cluster evolution in \S\ref{gravrad}, it is convenient to distinguish
three broad categories of dense stellar systems: (1) young populous
clusters, (2) globular clusters, and (3) galactic nuclei.
Table\,\ref{Tab:clusters} lists characteristic parameters for each.
The masses and virial radii of globular clusters are assumed to be
distributed as independent Gaussians with means and dispersions as
presented in the table; this assumption is supported by correlation
studies (Djorgovski \& Meylan 1994)\nocite{DM94}.  Table
\ref{Tab:clusters} also presents estimates of the parameters of
globular clusters at birth (bottom row), based on a recent
parameter-space survey of cluster initial conditions (Takahashi \&
Portegies Zwart 2000);\nocite{TPZ20} globular clusters which have
survived for a Hubble time have lost $\apgt 60$\% of their initial
mass and have expanded by about a factor of three.  We draw no
distinction between core-collapsed globular clusters (about 20\% of
the current population) and non-collapsed globulars---the present
dynamical state of a cluster has little bearing on how black-hole
binaries were formed and ejected during the first few Gyr of the
cluster's life.

\section{Production of gravitational radiation}\label{gravrad}
An approximate formula for the merger time of two stars due to the
emission of gravitational waves is given by Peters \& Mathews 1963):
\nocite{PM63}
\begin{equation}
 t_{\rm mrg} \approx 150\, {\rm Myr}\; 
	\left( {\msun \over \mbh} \right)^{3}
	\left( {a \over \rsun} \right)^4 (1-e^2)^{7/2} \;.
\label{Eq:tmrg}\end{equation}
The sixth column of Table\,\ref{Tab:clusters} lists the fraction of
black-hole binaries which merge within a Hubble time due to
gravitational radiation, assuming that the binary binding energies are
distributed flat in $\log E_b$ between $1000\,kT$ and $10000\,kT$,
that the eccentricities are thermal, independent of $E_b$, and that
the universe is 15\,Gyr old 
(Jha et al.\ 1999).\nocite{Jha99}
The final column of the table lists
the contribution to the total black-hole merger rate from each cluster
category.


\subsection{Merger rate in the local universe}
Given the black-hole merger rate corresponding to each category of
star cluster, we now estimate the total merger rate ${\cal R}$ per
unit volume.  Table\,\ref{Tab:galaxies} lists, for various tyes of
galaxies, the space densities and $S_N$, the specific number of
globular clusters per $M_v = -15$ magnitude (van den Bergh
1995):\nocite{vdBerg95}
\begin{equation}
	S_N = N_{GC} 10^{0.4(M_v + 15)}
\end{equation}
(where $N_{GC}$ is the total number of globular clusters in the galaxy
under consideration).  The values given for $S_N$ in Table
\ref{Tab:galaxies} are corrected for internal absorption; the absorbed
component is estimated from observations in the far infrared.  The
estimated number density of globular clusters in the universe is
\begin{equation}
	\phi_{GC}  = 8.4\,h^3\; {\rm Mpc^{-3}}.
\end{equation}
A conservative estimate of the merger rate of black-hole binaries
formed in globular clusters is obtained by assuming that globular
clusters in other galaxies have characteristics similar to those found
in our own.  The result is
\begin{equation}
	{\cal R}_{GC} = 5.4\times 10^{-8} h^3\; 
	                {\rm yr}^{-1}\,{\rm Mpc}^{-3}.
\end{equation}

Irregular galaxies, starburst galaxies, early type spirals and blue
elliptical galaxies all contribute to the formation of young populous
clusters.  In the absence of firm measurements of the numbers of young
populous clusters in other galaxies, we simply use the same values of
$S_N$ as for globular clusters.  The space density of such clusters is
then
$	\phi_{YPC}  = 3.5\,h^3 \; {\rm Mpc^{-3}},   $
and the black hole merger rate is
\begin{equation}
      {\cal R}_{YPC} = 2.1 \times 10^{-8} h^3\;
				{\rm yr}^{-1}\,{\rm Mpc}^{-3}.
\end{equation}
We find that galactic nuclei contribute negligibly to the total black
hole merger rate.

Based on the assumptions outlined above, our estimated total merger
rate per unit volume of black-hole binaries is
\begin{equation}
	{\cal R} = 7.5 \times 10^{-8} h^3\; 
	           {\rm yr}^{-1}\,{\rm Mpc}^{-3}.
\label{total-merger-rate}
\end{equation}
However, this may be a considerable underestimate of the true rate.
First, as already mentioned, our assumed number ($\sim10^{-4}N$) of
ejected black-hole binaries is quite conservative.  
Second, the observed population of globular clusters naturally
represents only those clusters that have survived until the present
day.  The study by Takahashi \& Portegies Zwart (2000)\nocite{TPZ20}
indicates that $\sim$ 50\% of globular clusters dissolve in the tidal
field of the parent galaxy within a few billion years of formation.
We have therefore underestimated the total number of globular
clusters, and hence the black-hole merger rate, by about a factor of
two.  Third, a very substantial underestimate stems from the
assumption that the masses and radii of present-day globular clusters
are representative of the initial population.  When estimated initial
parameters (Table\,\ref{Tab:clusters}, bottom row) are used, the total
merger rate increases by a further factor of six.  Taking all these
effects into account, we obtain a net black-hole merger rate of
\begin{equation}
	{\cal R} \sim 3 \times 10^{-7} h^3\; 
	           {\rm yr}^{-1}\,{\rm Mpc}^{-3}.
\label{best_guess}
\end{equation}
We note that this figure is significantly larger than the current best
estimates of the neutron-star merger rate.  

\subsection{LIGO observations}
%
The maximum distance within which LIGO-I can detect an inspiral event
is estimated to be
\begin{equation}  
	R_{\rm eff} = 18\,{\rm Mpc}\ 
			\left(\frac{M_{\rm chirp}}{\msun}\right)^{5/6}
\end{equation}
(K.~Thorne, private communication).  Here, the ``chirp'' mass for a
binary with component masses $m_1$ and $m_2$ is $M_{\rm chirp} = (m_1
m_2)^{3/5} / (m_1+m_2)^{1/5}$.  For neutron star inspiral, $m_1 = m_2
= 1.4\,\msun$, so $M_{\rm chirp} = 1.22\,\msun$, $R_{\rm eff} = 21$
Mpc, and we obtain the detection rate mentioned in the introduction.
For black-hole binaries with $m_1 = m_2 = \mbh = 10\,\msun$, we find
$M_{\rm chirp} = 8.71\,\msun$, $R_{\rm eff} = 109$ Mpc, and a LIGO-I
detection rate of about 1.7\,$h^3$ per year.  For $h \sim 0.65$ (Jha
1999),\nocite{Jha99} this results in about one detection event every
two years.  LIGO-II should become operational by 2007, and is expected
to have $R_{\rm eff}$ about ten times greater than LIGO-I, resulting
in a detection rate 1000 times higher, or about one event per day.

\section{Discussion}
Black-hole binaries ejected from galactic nuclei, the most massive
globular clusters (masses $\apgt 10^6\,\msun$), and globular clusters
which experience core collapse soon after formation, tend to be very
tightly bound, have high eccentricities and merge within a few million
years of ejection.  These mergers therefore trace the formation of
dense stellar systems with a delay of a few Gyr (the typical time
required to form and eject binaries), making these systems unlikely
candidates for LIGO detections, as the majority merged long ago.  This
effect may reduce the current merger rate by an order of magnitude,
although more sensitive future gravitational wave detectors may see
some of these early universe events.  We estimate that the most
massive globular clusters contribute about 90\% of the total black
hole merger rate.  However, while their black-hole binaries merge
promptly upon ejection, the longer relaxation times of these clusters
mean that binaries tend to be ejected much later than in lower mass
systems.  Consequently, we have retained these binaries in our final
merger rate estimate (Eq. \ref{best_guess}). But we note that
represents a significant source if uncertainty.

By the time the black hole binary is ejected it has experienced
$\sim40$--50 hard encounters with other black holes, as well as a
similar number of encounters with other stars or binaries.  During
each of these latter encounters, there is a small probability that a
low-mass star may collide with one of the black holes.  Such
collisions tend to soften the black hole binary somewhat (see
Portegies Zwart et al.~1999), but they are unlikely to delay ejection
significantly.  A collision between a main-sequence star and a black
hole may, however, lead to brief but intense X-ray phase.

Finally, we have assumed that the mass of a stellar black hole is
10\,\msun.  Increasing this mass to 18{\msun} decreases the expected
merger rate by about 50\%---higher mass black holes tend to have wider
orbits.  However, the larger chirp mass increases the signal to noise,
and the distance to which such a merger can be observed increases by
about 60\% and the overall detection rate on Earth increases by about
a factor of three.  For 6 {\msun} black holes, the detection rate
decreases by a similar factor.
For black-hole binaries with component masses $\apgt 12$\,\msun, the
first generation of detectors will be more sensitive to the merger
itself than to the inspiral phase that precedes it (Flanagan \& Hughes
1998)\nocite{FH98}.  Since the strongest signal is expected from
black-hole binaries with high-mass components, it is critically
important to improve our understanding of the merger waveform.  Even
for lower-mass black holes (with $m_{bh}\apgt 10\,\msun$), the
inspiral signal comes from an epoch when the holes are so close
together that the post-Newtonian expansions used to calculate the wave
forms are unreliable (Brady et al.~1998)\nocite{BCT98}.


\bigskip\noindent{\bf Acknowledgments} We thank Piet Hut, Jun Makino
and Kip Thorne for insightful comments on the manuscript.  This work
was supported by NASA through Hubble Fellowship grant HF-01112.01-98A
awarded (to SPZ) by the Space Telescope Science Institute, which is
operated by the Association of Universities for Research in Astronomy,
Inc., for NASA under contract NAS\, 5-26555, and by ATP grant
NAG5-6964 (to SLWM).  SPZ is grateful to Drexel University, Tokyo
University and the University of Amsterdam (under Spinoza grant 0-08
to Edward P.J. van den Heuvel) for their hospitality.  Calculations
are performed on the GRAPE-4 computers at Tokyo University and Drexel
University, and on the SGI/Cray Origin2000 supercomputer at Boston
University.

\vfill
\newpage

\begin{table}
\caption[]{Overview of selected parameters for young populous clusters
(Massey \& Hunter 1998)\nocite{MH98}, globular clusters (Djorgovski \&
Meylan, 1994)\nocite{DM94} and galactic nuclei (McGinn et
al.~1989)\nocite{McGinn89}.  The first three columns list the cluster
type, the total mass (in solar units) and the virial radius (in pc).
The total mass and virial radius are given as
distributions with a mean and the standard deviation around the mean.
The orbital separation (in solar units) for a 1000\,$kT$ binary
consisting of two 10\,\msun\, black holes is given in the fourth
column.  The fifth and sixth columns list the expected number of
black-hole binaries that are formed by the cluster and the fraction of
these binaries which merge within 12\,Gyr (allowing $\sim 3$\,Gyr for
the formation and ejection of the binaries and assuming a 15\,Gyr old
universe, Jha 1999\nocite{Jha99}). The contributions to the total
black hole merger rate per star clusters per year (MR) are given in
the final column.  The bottom row contains estimated parameters for
the zero-age population of globular clusters in the Galaxy, indicated
with $^\star$. 
}
\medskip\begin{tabular}{lcccrrl}  
cluster  & M           &\rvir   &$1000\,kT$&$N_{\rm b}$&$f_{\rm merge}$&MR\\ 
type     &[$\log$]     & [$\log$] &[\rsun]    &     & &[Myr$^{-1}$]\\ 
Populous & 4.5         &  -0.4    &   420     &  7.9& 7.7\%& 0.0061  \\ 
Globular & $5.5\pm0.5$ &$0.5\pm 0.3$& 315     &  150& 51 \%& 0.0064  \\ 
Nucleus  & $\sim 7$    &$\aplt 0$ &$\aplt 3.3$& 2500&100 \%& 0.21    \\ 
Globular$^\star$
         & $6.0\pm0.5$ &$0\pm 0.3$&    33     &  500& 92 \%& 0.038 \\ 
\end{tabular}
\label{Tab:clusters} 
\end{table}

\vfill
\newpage

\begin{table}
\caption[]{Galaxy morphology class, space densities, average absolute
magnitude (Heyl et al.~1997)\nocite{heyl97},
and the specific frequency of globular clusters $S_N$ (from van den
Bergh, 1995 and McLaughlin 1999)\nocite{McL99}\nocite{vdBerg95}.  The
final column gives the contribution to the total number density of
globular clusters.}
\medskip\begin{tabular}{lrlcr} 
Galaxy   & $\phi_{\rm GN}$&$M_v$          &$S_Nh^2$& GC space density\\
Type     &[$10^{-3}\,h$\,Mpc$^{-3}$]& &       & [$h^3$ Mpc$^{-3}$] \\  
E--S0    & 3.49   & -20.7 & 10       & 6.65 \\
Sab      & 2.19   & -20.0 &  7       & 1.53 \\
Sbc      & 2.80   & -19.4 &  1       & 0.16 \\
Scd      & 3.01   & -19.2 &  0.2     & 0.03 \\ 
Blue E   & 1.87   & -19.6 & 14       & 1.81 \\
Sdm/StarB& 0.50   & -19.0 &  0.5     & 0.01 \\  
%
\end{tabular}
\label{Tab:galaxies} 
\end{table}

\end{document}